# THE UNIFIED APPROACH FOR ORGANIZATIONAL NETWORK VULNERABILITY ASSESSMENT

[1]Mrs. Dhanamma Jagli, [2]Mrs.Rohini Temkar

[1,]Research Scholar, JNT University, Hyderabad.
[1, 2] Assistant Professor, Department of MCA
V.E.S. Institute of Technology,
University of Mumbai, India.

*ABSTRACT:*

*The present business network infrastructure is quickly varying with latest servers, services, connections, and ports added often, at times day by day, and with a uncontrollably inflow of laptops, storage media and wireless networks. With the increasing amount of vulnerabilities and exploits coupled with the recurrent evolution of IT infrastructure, organizations at present require more numerous vulnerability assessments. In this paper new approach the Unified process for Network vulnerability Assessments hereafter called as a unified NVA is proposed for network vulnerability assessment derived from Unified Software Development Process or Unified Process, it is a popular iterative and incremental software development process framework.*

## I.  INTRODUCTION

The Unified Modeling Language (UML) is a language of modelling that provides design notations that is hastily apt an effectively paradigm software design language. UML offer a variety of constructive capabilities to the software designer, together with multiple, consistent design views, and a semiformal semantics articulated as a UML Meta model, and an allied modelling language intended for expressing prescribed logic constraints on design fundamentals. The main aim of this approach is to consider the power of UML for modelling software architectures in the way in which the amount of existing software Architecture Description Languages (ADLs) to model architectures. This paper presents two strategies which supports architectural concerns provided by UML. One approach involves using UML "as is," and the other incorporates useful characteristics of existing ADLs as UML extensions. At the same time as it talks about the applicability, strengths, and weaknesses of the two strategies. The strategy is practically applied on three ADLs, which represent a broad example of present-day ADL capabilities. Solitary conclusion of our work is that UML still lacks in following to capture and exploiting definite architectural concerns. The significance has been recognized throughout the research and exercise of software architectures. In particular, UML lacks shortest support for modeling and exploiting architectural styles, open software connectors, local and global architectural constraints.

## II. LITERATURE REVIEW

Software architecture is a segment of software engineering focussed on developing huge and complex applications in a manner that reduces development expense, increases the more potential meant for cohesion among various members of a directly related product family, and facilitates growth, possibly at system runtime. To date, the software architecture research community has focused principally on analytic evaluation of architectural descriptions. Several researchers have moved towards to believe that the benefit of an explicit architectural is centre of attention. Software architecture must be provided by means of its own body of specification languages and analysis techniques. grown-up engineering disciplines are normally categorized by accepting logical standards to describe all related artifacts of their subject matter, such standards are not only facilitate practitioners to collaborate, but they also contribute to the enlargement of the entire discipline.

### 1. The Unified Process

The Unified Software Development Process or else Unified Process are an admired iterative and incremental software development process. The well-known and generally documented enrichment of the unified process is the Rational Unified Process (RUP).the outline of a classical project presenting the relative dimensions of the four phases of the unified process. The Unified Process is not plainly a process, but rather an extensible framework which should be adapted for specific organizations or projects. The Rational Unified Process is similar to a customizable framework. As a consequence it is frequently not possible to state whether a refinement of the process is derived from UP or else from RUP.so that the names be to be expected to use synonymously.

### 2. The Unified Process phases:

The Unified Process has 4 phases as shown in the Fig 1.

1)    Inception:  Requirements capture and analysis
2)    Elaboration:  System and class-level design
3)    Construction:  Implementation and testing
4)    Transition:  Deployment

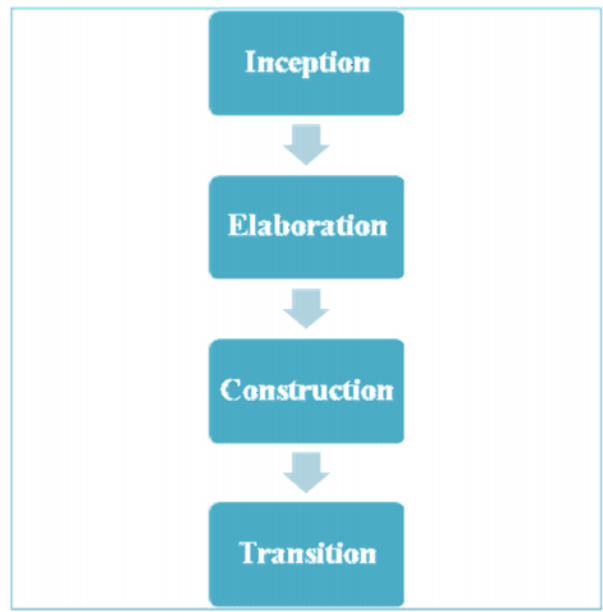

Figure 1: The Four Phases of Unified process

### a) Inception Phase

Inception is the least phase in the project development, and ideally it should be quite small. If the inception phase is extended then it may be a hint of excessive up-front requirement, which is unlike to the strength of the unified process. The inception phase is a foundation stage that attempts to answer the following questions:

- ✓ What is the purpose of the project?
- ✓ Is the project realistic (e.g. technologically, financially, with current personnel)?
- ✓ Should we buy the project, or build it?
- ✓ Will it be developed now, or built from a previously existing project?
- ✓ What are the estimated costs?
- ✓ Should we proceed with the project?

The following are characteristic goals for the inception phase:

- ✓ launch a explanation or else business case for the project
- ✓ Launch the scope of project and boundary settings of project.
- ✓ summarize the use cases and important key requirements that will drive the design tradeoffs
- ✓ sketch out one otherwise more applicant architectures
- ✓ recognize the risks
- ✓ Prepare a preliminary schedule for the project and cost estimate

The purposes of lifecycle milestone results are the end of the inception phase. Develop an estimated visualization of the system, make the business case, define the scope, and produce irregular estimation cost and schedule. This phase chiefly deals by way of project planning and

International Journal of Software Engineering & Applications (IJSEA), Vol.4, No.5, September 2013

project management this includes project plans and Gantt charts, budgets. Since the UP is incremental, the inception phase begins with some activities and produces some initial artifacts, such as a UML diagram or set of diagrams, that is produced during a phase. However, these activities will be continued in later phases and the artifacts modified, and new artifacts added some inception artefacts include:

- ✓ Business case: Describes high-level goals and constraints, in business terminology
- ✓ Use-Case model: Describes functional requirements, and related non-functional req.
- ✓ Risk management plan: Describes the business, resource, technical, and schedule risks and how to minimize these risks

### b) Elaboration Phase

During the Elaboration phase the project team is expected to confine a healthy majority of the system requirements. Even though, the most important goals of Elaboration phase are to deal with recognized risk factors to begin validating the system architecture. familiar processes undertaken in this phase consists the  formation of conceptual diagrams ,use case diagrams, class diagrams with simple basic notation and package diagrams or architectural diagrams. The architecture is validated first and foremost through  the execution of an Executable Architecture Baseline.  This is a partial implementation of the system which includes core, the most architecturally significant component. It has built into a sequence of tiny, timeboxed iterations. Besides the ending of the Elaboration phase the system architecture have to stabilize and the executable architecture baseline ought to exhibit that the architecture determine the supporting the main system functionality and demonstrate the accurate actions in terms of  performance, scalability and price tag. The final Elaboration phase deliverable is a sketch including expenditure and schedule estimates used for the Construction phase. At the end the plan should be correct and convincing; since it should be based on the Elaboration phase occurrence and significant risk factors should have been addressed throughout the Elaboration phase. The function of the inception phase is to understand the difficulty. Often 1-3 iterations are required for Elaboration.

The most important purpose of the Elaboration phase is to start to understand how the software will resolve particular problem. In other words, this is where much of the architecture and design of the software takes place. After Elaboration, project risks are essentially eliminated. The user interface has been approved by customers and managers. Technically difficult software components were implemented or proof-of-concept code was created to prove it was possible. Cost estimates are finalized, so budgets can be approved. Preliminary user manuals have been created and analyzed. Analysis, architecture and design well underway after Elaboration.E.g. Use cases should be about 80% complete.

The Elaboration phase involves:

- ✓ Continuing work on the use-case model.
- ✓ In particular, during the elaboration phase is when activity diagrams might be added to describe how use cases are to operate.
- ✓ Diagrams showing how participants (objects and actors) interact during the problem solving process (system sequence diagrams).

- ✓ Prototypes and proof-of-concepts: Some of the more difficult system aspects to accomplish are solved with GUI prototypes, sample database contents, and sample OOPL code for difficult algorithms.

**c)    Construction Phase**

Built on the foundation laid in Elaboration. System characteristics are implemented in a sequence of iterations. Each iteration of Unified Process results in an executable release of the software. It is accepted to put in writing complete text use cases for the duration of the construction phase and each one becomes the switch on to new-fangled iteration. Widespread UML (Unified Modelling Language) diagrams are used throughout this phase and encompass, Collaboration, State Transition, Activity, Sequence and Interaction Overview diagrams.

The construction phase involves iterative enhancement to previously created artifacts:

- ✓ Domain model
- ✓ Design model
- ✓ Implementation

Obviously, the implementation is the most significant body of work enhanced during construction.

**c)    Transition Phase**

The concluding project phase is Transition phase. In this phase the system is deployed to the intended users. The response received from an initial release or else initial releases may possibly give results in additional refinements to be integrated more than the course of more than a few Transition phase iterations. The Transition phase also includes system conversions and user training. During this phase, the software produced at the end of the construction phase is deployed this could involve:

- ✓ Rigorous complete system testing
- ✓ Installation programs being purchased or developed
- ✓ Software media being developed
- ✓ Solutions for support/user training being implemented
- ✓ Best testing under varied deployment environments
- ✓ Verifying that the system meets acceptance criteria

**3.    The Unified Process Is Iterative and Incremental**

Developing a business software product is a huge accountability that may prolong over more than a few months to possibly a year or more. It is n iteration that n iteration that practical to divide the work into smaller slices or mini-projects .Each mini-project is an iteration that grades in an increment. Iteration refers to stepladder in the workflow, and increments to growth in product. To be most effective, the iterations must be controlled; that is they must be selected and carried out in a planned way. This is why they are mini-projects. Developers stand for the choosing what is to be implemented within iteration ahead two factors. Primary, the iteration deals with a set of uses cases that collectively enlarge the usability of the product as developed until now. Second,

International Journal of Software Engineering & Applications (IJSEA), Vol.4, No.5, September 2013

the iteration deals with the most important risks. Successive iterations build on the development artefacts from the state at which they were absent at the end of the preceding iteration. It is a mini-project, so from the use cases it continues through the consequent development work-analysis, design, implementation, and test-that realizes in the form of executable code the use cases being developed in the iteration. Of course, an increment is not necessarily additive. Especially in the early phases of the life cycle, developers may be replacing a superficial design with a more detailed or sophisticated one. In later phases increments are typically additive. Inside every iteration, the developers recognize and state the relevant use cases, create a design by using the selected architecture as a channel, put into practice the design in components and authenticate that the components convince the use cases. Condition iteration meets its goals in addition to development proceeds with the after that iteration. Once iteration does not meet up its goals, the developers have to come flooding back to their previous decision and attempt an innovative approach. To achieve the greatest economy in development, a project team will try to select only the iterations required to reach the project goal. It will try to sequence the iterations in a logical order. A successful project will proceed along a straight course with only small deviations from the course the developers initially planned. Of course, to the extent that unforeseen problems add iterations or alter the sequence of iterations, the development process will take more effort and time. Minimizing unforeseen problems is one of the goals of risk reduction. There are many benefits to a controlled iterative process:

o   Controlled iteration decreases the cost risk to the expenditures on a single increment. If the developers need to repeat the iteration, the organization loses only the misdirected effort of one iteration, not the value of the entire product.
o   Controlled iteration reduces the risk of not getting the product to market on the planned schedule. By identifying risks early in development, the time spent resolving them occurs early in the schedule when people are less rushed than they are late in the schedule. In the "traditional" approach, where difficult problems are first revealed by system test, the time necessary to resolve them usually exceeds the time remaining in the schedule and nearly always forces a delay of delivery.
o   Controlled iteration speeds awake the cadence of the total development effort since developers effort more efficient towards results in comprehensible, undersized centre of attention relatively rather than an ever-sliding and long, schedule.
o   Controlled iteration acknowledges a truth often uncared for that user requirement and the equivalent requirements cannot be completely defined forthright. They are typically refined in successive iterations. This mode of operation makes it easier to adapt to changing requirements.

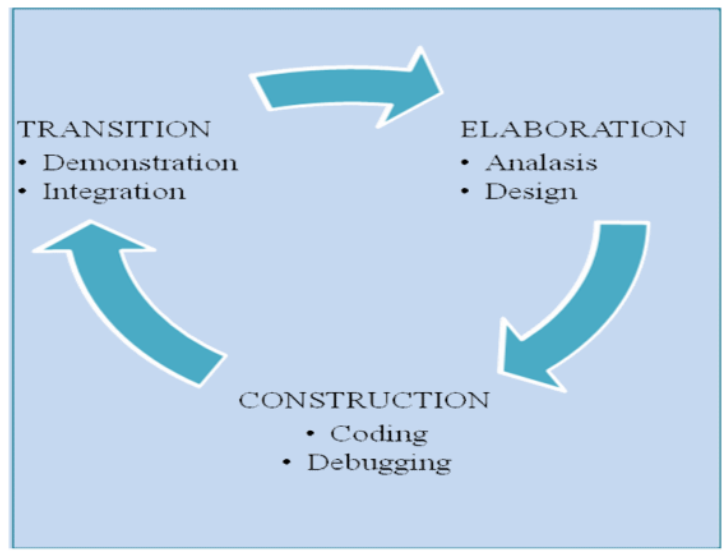

Figure 2: The Iterative Unified process

The victorious commercialization of numerous applications of wireless networks depends on the guarantee of the confidentiality, availability and integrity of the data communicate throughout the network. The confidentiality is defined as the ability to keep on maintains data secret from each and every one excluding a set of authorized entities. The Integrity is defined as the capability to confirm that data has not been maliciously or else by accidentally altered while in transit. Recent researchers had demonstrated that these important properties can be powerfully compromised by physically capturing network nodes and extracting cryptographic secret keys from their remembrance such node capture attack are very much possible due to the potential unattended procedure of wireless nodes and the prohibitive expenditure of tamper-proof hardware in transportable devices. By using the cryptographic keys recovered in a node capture attack, an adversary can conciliation the confidentiality and integrity of any messages protected using the compromised keys.

## 4. Vulnerability

It is the intersection of three elements: a system susceptibility or flaw, attacker access to the flaw, and attacker capability to exploit the flaw. [1] To exploit vulnerability, an assailant must have at least one suitable tool or method that can attach to a system weak spot. During this frame, vulnerability is also known as the attack facade.

### a)     Vulnerability Types

Vulnerabilities are classified according to the asset class they are related to:

1. Hardware
    A. Susceptibility to humidity
    B. Susceptibility to dust
    C. Susceptibility to soiling

43

    D.  Susceptibility to unprotected storage
2. Software
    A.  Insufficient testing
    B.  Lack of audit trail
3. Network
    A.  Unprotected communication lines
    B.  Insecure network architecture
4. Personnel
    A.  Inadequate Recruiting Process
    B.  Inadequate Security Awareness
5. Site
    A.  Area Subject To Flood
    B.  Unreliable Power Source
6. Organizational
    A.  Lack of Regular Audits
    B.  Lack of Continuity Plans
    C.  Lack of Security

**b) Identifying and removing vulnerabilities**

Many software tools exist that can assist in the discovery (and sometimes removal) of vulnerabilities in a computer system. Although these security supporting tools can supply an auditor with a high-quality overview of likely possible vulnerabilities present, they can not restore human being verdict. Relying exclusively on scanners will give up bogus positives and a limited-scope outlook of the nuisance present in the system. Vulnerabilities have been initiate in every main operating system including Windows, Mac OS, various forms of UNIX and Linux, OpenVMS, and others. The simply way to diminish the opening of a vulnerability being used against a system is through steady observation, including cautious system safeguarding e.g. applying software patches, most excellent practices in use e.g. the use of firewalls, access controls and auditing both throughout development and the deployment lifecycle.

## III. THE PROPOSED MODEL

The Unified NVA is the Unified process for Network vulnerability assessment is a well-organized approach for network vulnerabilities findings that can be exploited by a determined intruder to add access to or shut down a network. Network vulnerability is a situation, a weakness of or an absence of security methods, technical controls, or physical controls that could be oppressed by a great threat

### 1. The Unified NVA Phases
The Unified NVA has 4 phases as shown in the Fig 3.
1) Initiation
2) Expansion
3) Erection
4) Evolution

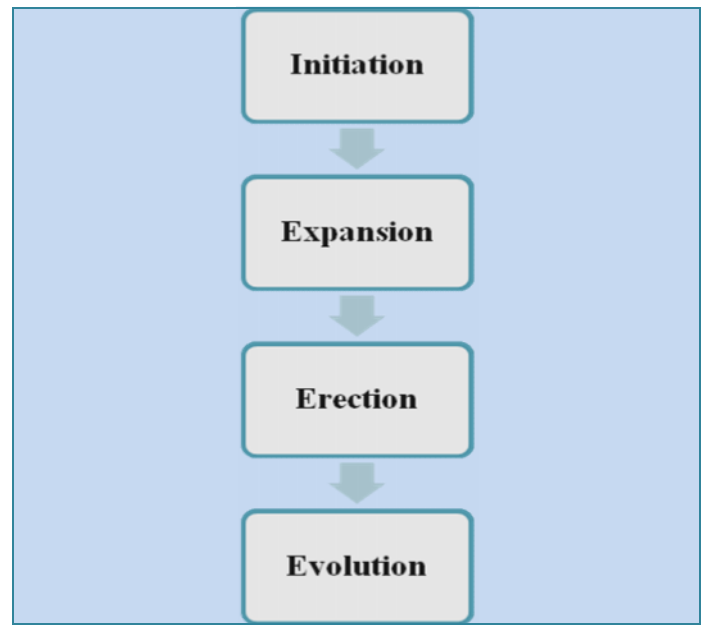

Figure 3: The Unified NVA Phases

**a)      Initiation Phase**

Risk analysis is the initiation for NVA. Assessing risk is a process and as such, is something that must be periodically repeated. It's really not much different from the automated patch-management tools are probably using. True security requires ongoing effort. There is never a wrong time to assess risk and examine network vulnerabilities. There are four ways in which can respond to risk: avoidance, transference, mitigation, and acceptance:

**b)      Expansion Phase**

Policy assessments will call this a level I assessment. It is a top-down gaze at procedures, guidelines and the organization's policies. This type of vulnerability assessment does not contain any hands-on testing the purpose of a top-down policy assessment is to answer three questions:

- ✓   Do the applicable policies exist?
- ✓   Are they being followed?
- ✓   Is there content sufficient to guard against potential risk?

**c)      Erection Phase**

Rolling out new policy may seem to be no more difficult than dictating the date of expected compliance, but that is not the case. Consider a major change to the authentication and authorization policy.  Mandate that everyone must change to complex passwords on the first day of the next month. So, on Monday morning, all employees attempt to change passwords and many experience problems. The result is that the help desk is flooded with calls and many individuals experience an unproductive morning, waiting to log in and begin work. Policies should be implemented in such a way that the change is gradual, staged, or piloted. Many individuals already have the belief that security policies inhibit work and slow things down, so want to make sure that any change make does not contribute to that sentiment.

### d)      Evolution Phase

On one occasion the threats and vulnerabilities have been evaluated, design the penetration testing to deal with the risks recognized all over the environment. . The penetration testing should be suitable for the complexity and size of an organization. Penetration tests different from assessments and evaluations, penetration tests are adversarial in environment.  Will refer to penetration tests as level III assessments. All locations of sensitive data, all key applications that store, process, or transmit such data, all key network connections, and all key access points should be included. The penetration testing should try to exploit security vulnerabilities and weaknesses throughout the environment, attempt to penetrate both at the network stage and key applications. The goal of penetration testing is to resolve if unauthorized access to key systems and accounts can be achieved. If access is achieved, the vulnerability ought to be corrected and the penetration testing re-performed in anticipation of the test is spotless and no longer allows unauthorized access or other malicious actions. These measures classically obtain on an adversarial responsibility and give the impression of being to see what the stranger can access and control inside the organization. Penetration tests are much less concerned with policies and procedures and are more focused on finding "low-hanging fruit" and seeing what a hacker can negotiate on this network. We almost always recommend that organizations complete an assessment and evaluation before beginning a penetration test, because a company with sufficient policies and procedures can't implement real security without documented controls. The general NVA life cycle as shown in the below Fig 4.

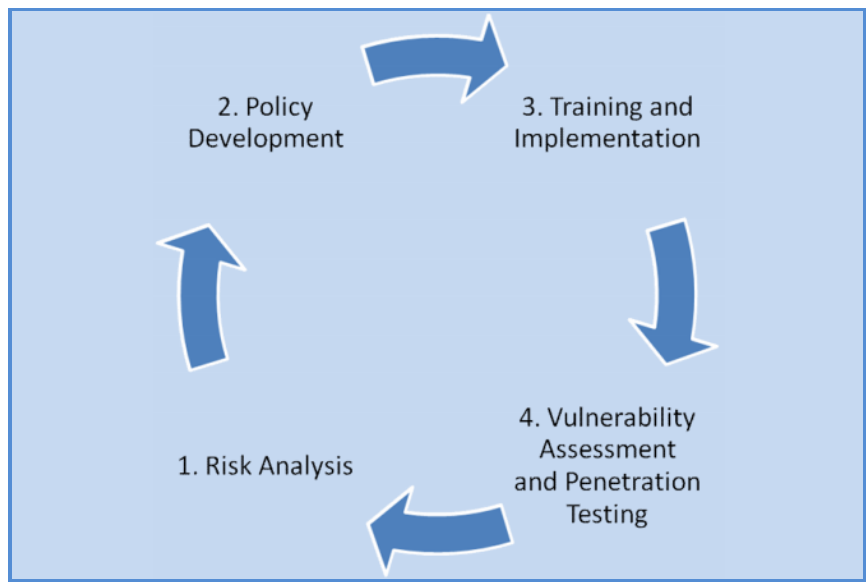

Figure 4: The NVA Life Cycle

International Journal of Software Engineering & Applications (IJSEA), Vol.4, No.5, September 2013

## IV. PHASES OF SYSTEM

Organizational network vulnerability assessment is the cyclical practice of identifying, classifying, remediating, and mitigating vulnerabilities this practice usually refers to network vulnerabilities in computing systems of any organizations. Successive iterations build on the development additive from the state at which they were left at the end of the previous iteration as shown in the below Fig 5.

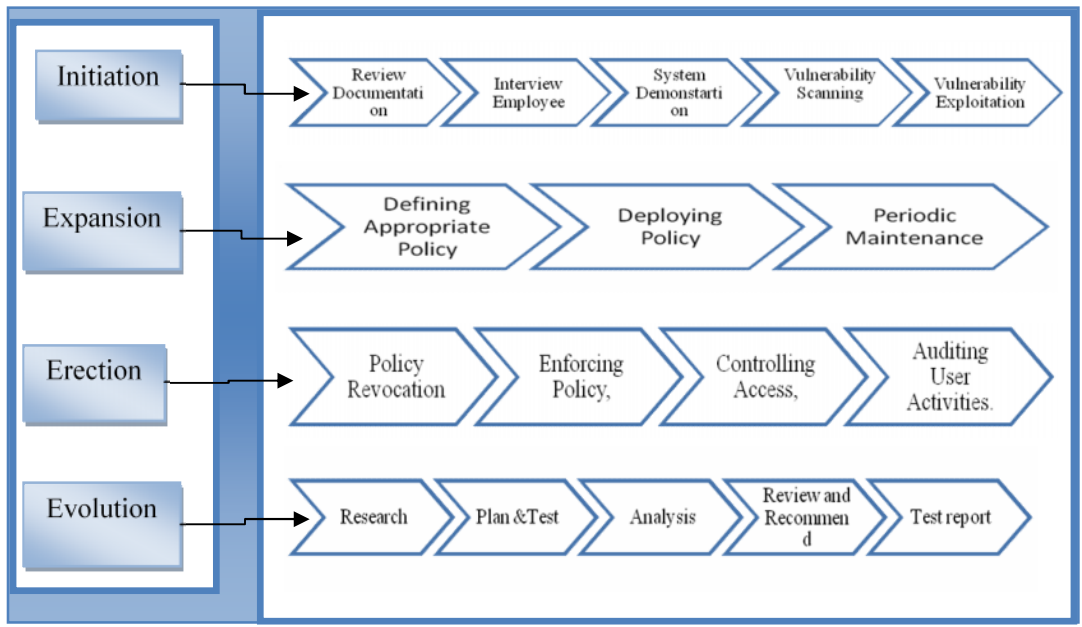

Figure 6: The Unified NVA Work flow

Network Vulnerability Assessor base the selection of what is to be done in iteration upon two factors. First, the iteration deals with a group of policies that together extend the usability of the network resources. Second, the iteration deals with the most important awareness or training to users.

## V. CONCLUSION AND FURTHER ENHANCEMENT

The unified network vulnerability approach is vault model which is proposed in this paper can be implemented as systematic approach provided. And it will be helpful for finding network vulnerabilities in any organization infrastructure. This model as an iterative process will help to increase efficiency and reduce the vulnerabities in the network resources and protect network assets from attackers.

### REFERENCES

[1] "Modeling Software Architectures in the Unified Modeling Language", by NENAD MEDVIDOVIC University of Southern California DAVID S. ROSENBLUM and DAVID F. REDMILES University of California, Irvine and JASON E. ROBBINS CollabNet, Inc.

[2] The text book, "Inside Network Security Assessment: Guarding your IT Infrastructure", by By Michael Gregg, David Kim.
[3] The text book,"The Unified Software Development Process ", By Ivar Jacobson, Grady Booch, James Rumbaugh, Pearson Education.
[4] K. Jain, "Security based on network topology against the wiretapping attack," IEEE Wireless Communication, vol. 11, no. 1, pp. 68–71, Feb.2004.
[5] D. Liu, P. Ning, and R. Li, "Establishing pairwise keys in distributed sensor networks," ACM Trans. Information and System Security, vol. 8,no. 1, pp. 41–77, Feb. 2005.
[6] W. A. Arbaugh, W. L. Fithen, and J. McHugh. Windows of Vulnerability: a Case Study Analysis. IEEE Computer, 2000.
[7] Agarwal, R., & Sinha, A. P. (2003). Object-Oriented Modeling with UML: A Study of Developers' Perceptions. Communications of the ACM, 46(9), 248-256.
[8] Atkinson, C., & Kühne, T. (2002). Rearchitecting the UML Infrastructure. ACM Transactions on Modeling and Computer Simulation, 12(4), 290-321.
[9] Barbier, F., Henderson-Sellers, B., Parc-Lacayrelle, A. L., & Bruel, J.-M. (2003). Formalization of the Whole-Part Relationship in the Unified Modeling Language. IEEE Transactions on Software Engineering, 29(5), 459-470.

**Authors**

**Mrs.Dhanamma Jagli** is an Assistance professor in V.E.S Institute of Technology, Mumbai, currently Pursuing Ph.D in Computer Science and Engineering and received M.Tech in Information Technology from Jawaharlal Nehru Technological University, Hyderabad. She has around 10 years teaching experience at the postgraduate and under graduate level. She had published and presented research papers in referred international journals and various national and international conferences. Her areas of research interest are Data Mining, Cloud Computing, Software Engineering, Data base Systems and Embedded Real time systems. She has been associated with Indian Society of Technical Education (ISTE) as a life member
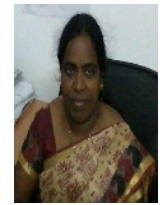

**Mrs.Rohini Temkar** is an Assistance professor in V.E.S's  Institute of Technology,Mumbai. She has completed her M.E. in Computer Engineering from University of Mumbai. She has around 11 years of teaching experience at the post graduate and under graduate level. She has published and presented technical papers in various international journals and conferences. Her areas of research interest are Object Oriented Modeling and Development, Web Development, Information Security, and Cloud Computing. She has also been associated with Indian Society of Technical Education (ISTE) as a life member.
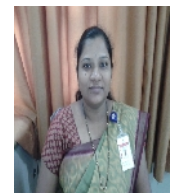